# Identification and predictability of the large-scale synoptic drivers contributing to the Kerala floods using multivariate feature-based analysis

Marion P. Mittermaier

*Global Model Evaluation & Development,*

*Foundation Science, Met Office, United Kingdom*

February 2025

This work and its contributors was conducted through the Weather and Climate Science for Service Partnership (WCSSP) India, a collaborative initiative between the Met Office, supported by the UK Department for Science, Innovation & Technology (DSIT) and the Indian Ministry of Earth Sciences (MoES).

Correspondence: marion.mittermaier@metoffice.gov.uk; FitzRoy Road, Exeter, EX1 3PB, UK

**Abstract**

Parts of Kerala were hit by devastating floods three years in a row. This sequence was historically without precedent. This paper uses a multivariate object-based approach to examine the global forecasts of the Met Office Unified Model for the 2018, 2019 and 2020 monsoon seasons to identify and understand the large-scale synoptic drivers that led to the floods. Identifying similarities between synoptic drivers behind these flooding events was a key objective, as was understanding whether one could enhance predictability by using a multivariate approach for post-processing forecast output, using variables other than just precipitation, which is often inherently less predictable on its own. To this end event identification focused on the analyses first, and then on a day 5 forecast.

The study found that the multivariate version of the Method for Object-based Diagnostic Evaluation (MvMODE) was able to successfully identify sequences of days in all three seasons corresponding to the event dates, which in combination, had flood-producing potential. This was achieved in both the analyses and in the 5-day forecasts, proving that a) the events share common synoptic drivers and b) there is inherent predictability which can be tapped into. The 5-day forecasts matched the analysed objects on each occasion, proving that the underlying large-scale drivers may be predictable into the medium-range.

Based on the results, the paper proposes a conceptual synoptic pattern evolution which can help identify such events in future. The synoptic pattern has some similarities to atmospheric rivers in the mid-latitudes.

**Significance statement**

The purpose of this work was to use a multivariate diagnostic verification method to examine the large-scale synoptic drivers that combined to create the 2018, 2019 and 2020 Kerala flooding events, and whether there are any similarities between these three events. It was found that the setup for the three years is similar and predictable in operational day 5 forecasts. Future work will focus on how far the predictability might extend into the medium range and beyond.

## 1. Introduction

The Kerala region of India was plagued by devastating flooding in three successive years between 2018 and 2020. Some or all these flood events have been studied in recent years (e.g., Joseph et al., 2020; Vijaykumar et al., 2021; Walia et al., 2023), covering topics as diverse as the impact of the changing climate to community resilience to disasters. On 16 August 2018 severe flooding affected the state and close to 500 people lost their lives and one million people were evacuated. The floods were exacerbated by reservoir storage already being near capacity due to persistent above-normal monsoon rains in the weeks preceding the event. Heavy rain was recorded between 8-10 August, and again between 12—18 August with 140 mm on August 15$^{th}$. A year later around August 8$^{th}$, Kerala again experienced severe flooding, with around 100 fatalities and thousands of people evacuated. Heavy rains persisted between 6—11 August with one day reporting more than 150 mm. Another year later, around August 7$^{th}$, Kerala was again severely affected by flooding with over 100 fatalities. All these events can be attributed to heavy monsoon rains, but what, if anything, did the events have synoptically in common that might have aided predictability? Can a multivariate approach identify the large-scale drivers that led to the sequence of events as they unfolded?

Kerala as a region is bounded by the Arabian Sea to the west, and the Western Ghats to the east. The coastline is roughly perpendicular to a low-level westerly jet which develops once the monsoon rains have started. The Western Ghats is a mountain range that runs all along the west coast of peninsular India, varying in height between 900—2600 m, with an average height of around 1600 m. The coastline and the proximity of the mountains to the coast provide the mechanisms for additional orographic enhancement. Given their height the Western Ghats present a barrier to perpendicular low-level flows up to at least 850 hPa, such that forced uplift contributes to rainfall accumulations in this region.

Atmospheric Rivers (ARs) are defined as long narrow channels of very moist air, only around 300 miles wide associated with a low-level jet stream ahead of the cold front of an extratropical cyclone. They are important poleward transporters of moisture, and can lead to heavy rain or snow, exacerbated by the presence of complex terrain which acts to enhance the precipitation further (Newell et al., 1992; Zhu & Newell, 1998; Ralph et al., 2004). In the last decade there is an increase in the frequency for extreme rainfall events over this region, and these extreme rainfall events are examined through the lens of being AR-like features. The hypothesis that is proposed here is that these floods were caused by *quasi*-atmospheric river-like behaviour, given the region of interest is not in the mid-latitudes, nor does it involve extratropical cyclones. Nevertheless, it will be shown that an analogous setup, consisting of a low-level moisture source, with a low-level jet exists, with the potential for convection to be enhanced by the presence of a monsoon depression.

Feature-based diagnostic evaluation (Davis et al., 2006; Nachamkin, 2004; Nachamkin et al., 2005; Rife & Davis, 2005) was borne out of a need for objective event-based analysis whereby the forecast performance is only viewed through the lens of that type of event. This way other aspects of the forecast, considered irrelevant to the event, are excluded. At times we use the term “perfect storm” when all the ingredients for a high-impact event come together, demonstrating that we inherently know that the events we see are due to a multivariate interaction. Here a multivariate feature-based approach is exploited to understand 1) how various atmospheric variables interact to lead to an observed sequence of events, and 2) whether a medium-range forecast can capture this. A brief description of the data used is provided in Section 2. In section 3 the methodology is described. Section 3 describes the data and Section 4 contains the results. Conclusions and a proposed life cycle of such events is presented in Section 5.

## 2. Data

The forecasts examined in this study are from the operational global version (GM) of the Met Office Unified Model (UM). The UM is used across space and time scales, and for both weather and climate applications. The UM is a nonhydrostatic, semi-implicit, semi-Lagrangian grid point model. At the time the GM was based on the Global Atmosphere (GA) 7 version of global deterministic configuration, which uses 70 terrain-following vertical levels. GA7 is based on the parameterisations described in Walters et al. (2019). The GM uses a hybrid data assimilation (DA) scheme (Clayton et al., 2012) and has a nominal horizontal resolution of ~10 km in the mid-latitudes but in the Tropics the 0.11° grid resolution is closer to 12 km in the west-east direction. GM forecasts are available out to 7 days (t+168h). Here only the day 5 (t+120h) forecasts initialised at 00 UTC are used to explore what the model is capable of in the medium range.

Whilst the t+0h analysis is utilised for evaluating variables such as winds and humidity, the Global Precipitation Measurement (GPM) Integrated Multi-satellitE Retrievals (IMERG, Hou et al., 2014; Huffman et al., 2019; Skofronick-Jackson et al., 2017) version 06B final products are used for precipitation. These data are available at a horizontal resolution of 0.1° x 0.1° (or ~10 km) with 30 minutes temporal resolution. Daily GPM accumulations were created and interpolated onto the GM grid. Many regional evaluation studies of GPM products exist. Prakash & Srinivasan (2021) provide an in-depth overview of GPM product performance over India. GPM data has also been extensively for model evaluation over the region (e.g., (Kolusu et al., 2023; Mittermaier et al., 2024).

## 3. Methodology

MvMODE is the multivariate extension of the Method for Object-based Diagnostic Evaluation (MODE, Davis et al., 2006). It is available via a comprehensive library of tools known as the Model Evaluation Tools (MET, Brown et al., 2021), which are wrapped in python configuration files collectively known as METplus (e.g., Win-Gildenmeister et al., 2021). The code is open source and available on github (https://github.com/DTCenter). The code base is curated by the National Center for Atmospheric Research (NCAR)'s Developmental Testbed Center (DTC) and is subject to regular updates.

MODE enables features of interest to be identified in a field. One has the option of smoothing the field before feature identification (referred to as applying a convolution radius). In this work a convolution radius of 5 grid lengths was used to smooth all fields. Beyond this step, MODE provides the capability of merging identified objects or features to create larger clusters in the same field. It also provides the capability to match features between two sets of fields so they can be analysed as pairs. This is achieved via a fuzzy logic dictionary, which chooses a range of object attributes to identify potential matches and compute an interest score to determine the goodness of a match. Only matches with interest scores greater than 0.7 are valid matches. This allows for a matched forecast-analysis pair analysis to explore how well the forecast performs, rather than just analysing the feature attributes in either field.

The multivariate logic added to MODE makes it possible to find the intersection between features identified in several different variables to create what is termed a “super” object. As before the super objects can be repopulated with any variable of interest to explore and compare the event (feature) that has been identified, in the forecast and observed (analysis) fields. As with univariate MODE, the super objects can then be paired to evaluate how well the forecast performs in predicting a certain type of event.

The objective is finding the overlaps between the objects obtained from the different (univariate) variables. Thus, unlike univariate MODE, the process of selecting thresholds for each of the components for the multivariate case is different. For univariate MODE, the aim is (usually) to be highly targeted in identifying the events of interest, whilst also minimising the number of objects. This approach does not work well when seeking the coincidence between features in different fields, and much lower thresholds (than one might expect) are required to aid in the process of finding the overlaps. In short, being too specific tends to hinder multivariate feature identification.

To help inform on the choice of thresholds, the relationships between the variables of interest were examined first, using another METplus tool called GridDiag, which enables the construction of bivariate or joint histograms.

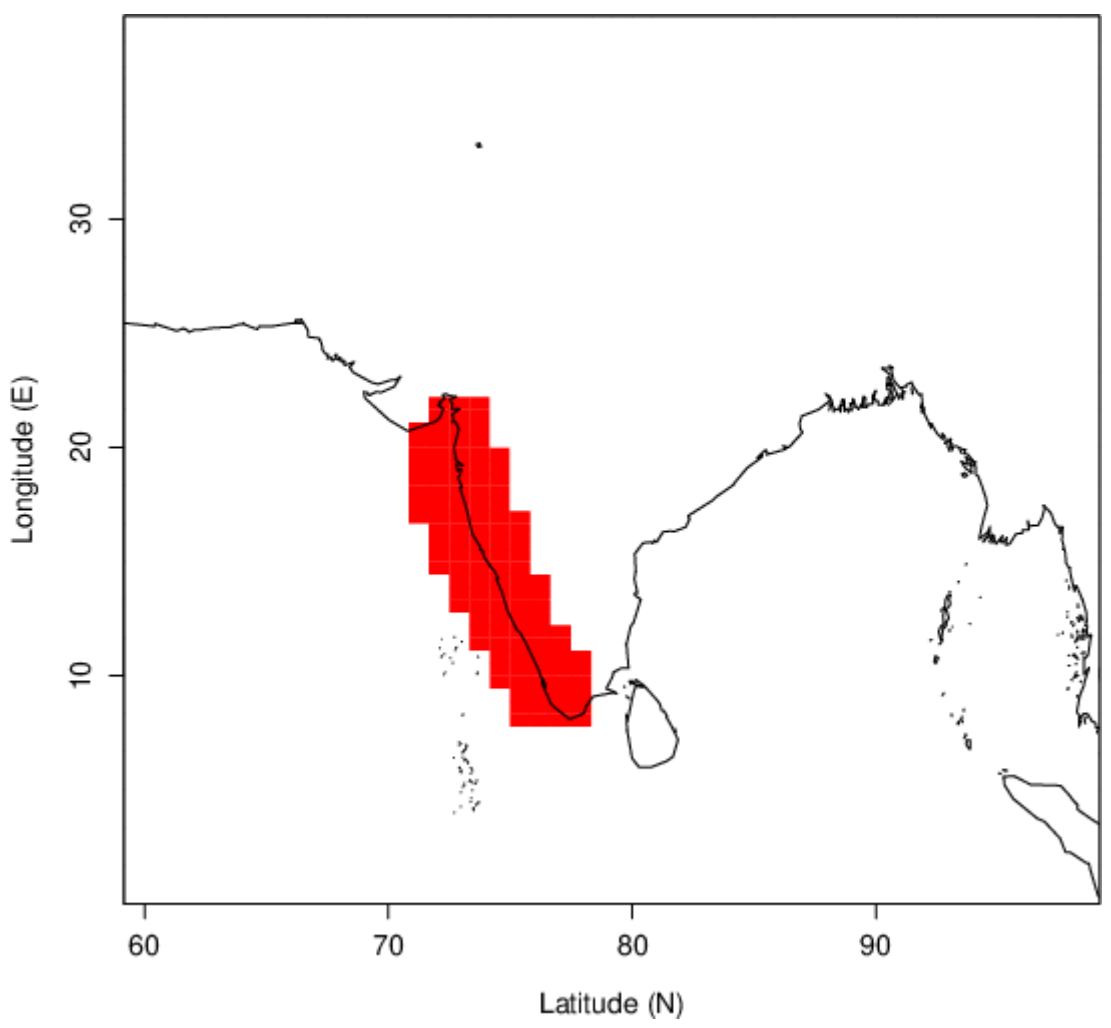

*Figure 1 Regional mask used for deriving suitable thresholds for feature identification.*

Figure 1 shows the region over which the bivariate distributions (defining the relationships between atmospheric variables at variables levels) were examined. To help with identifying the events, the behaviour of the variables of interest were compared over the time window 1—11 August 2019, which includes the days immediately before and during the 2019 flood producing rains. Values in the global model analyses (t+0h) and the GPM daily accumulations (PPN24), both regridded (area-weighted) to a coarser ~40 km grid, were analysed over this region. GridDiag examines each grid point in the two fields that fall within the area and allocates the value to a bin in a two-dimensional histogram with predefined bins to yield the joint distribution of how the values in the two fields vary together. These joint histograms (normalised and presented on a log10 scale) are plotted in Figure 2. Panels (a) and (b) show the relationship between the UWIND and VWIND at 850 hPa (a proxy for the low-level westerly jet and the meridional wind components) with the daily accumulations. These show a strong link between UWIND at 850 hPa (UWIND@850) speeds over 15 m $s^{-1}$ being associated with the largest totals. On the other hand, the VWIND component is close to 0 but potentially favouring something slightly north of west. Panels (c) and (d) provide a sense of the relationship between the relative humidity (RH) at 700 hPa (RH@700) and with the zonal and meridional wind components. The correspondence is strongest with RH higher than 80% and wind speeds as denoted in (a) and (b). Finally (e) and (f) show the relationship between RH at 850hPa and RH at 700 hPa with daily precipitation. Unsurprisingly there is a strong indication that totals greater than 50 mm $d^{-1}$ are far more likely when the humidity is at least 70—80%. The results are broadly similar for the two levels but given that moisture in depth is potentially important from an orographic enhancement perspective, RH@700 was preferred over RH@850.

Given this analysis of distributions, additional sensitivity tests were needed to settle on the final set of thresholds that would (reliably) provide a set of events to analyse. The remainder of the analysis that follows here is based on MvMODE configured with the following logic, variables, and thresholds: "UWIND@850 > 15 m $s^{-1}$ && RH@700 > 75% && PPN24 > 25 mm". As discussed earlier, pushing the thresholds higher had a negative effect on finding overlaps. Using VWIND at 850 hPa (VWIND@850) proved to be problematic, as a threshold of VWIND@850>0 was too non-specific. Though not shown, the distributions for other lead times, out to day 7, were remarkably similar, suggesting that the bivariate relationships explored using GridDiag are robust.

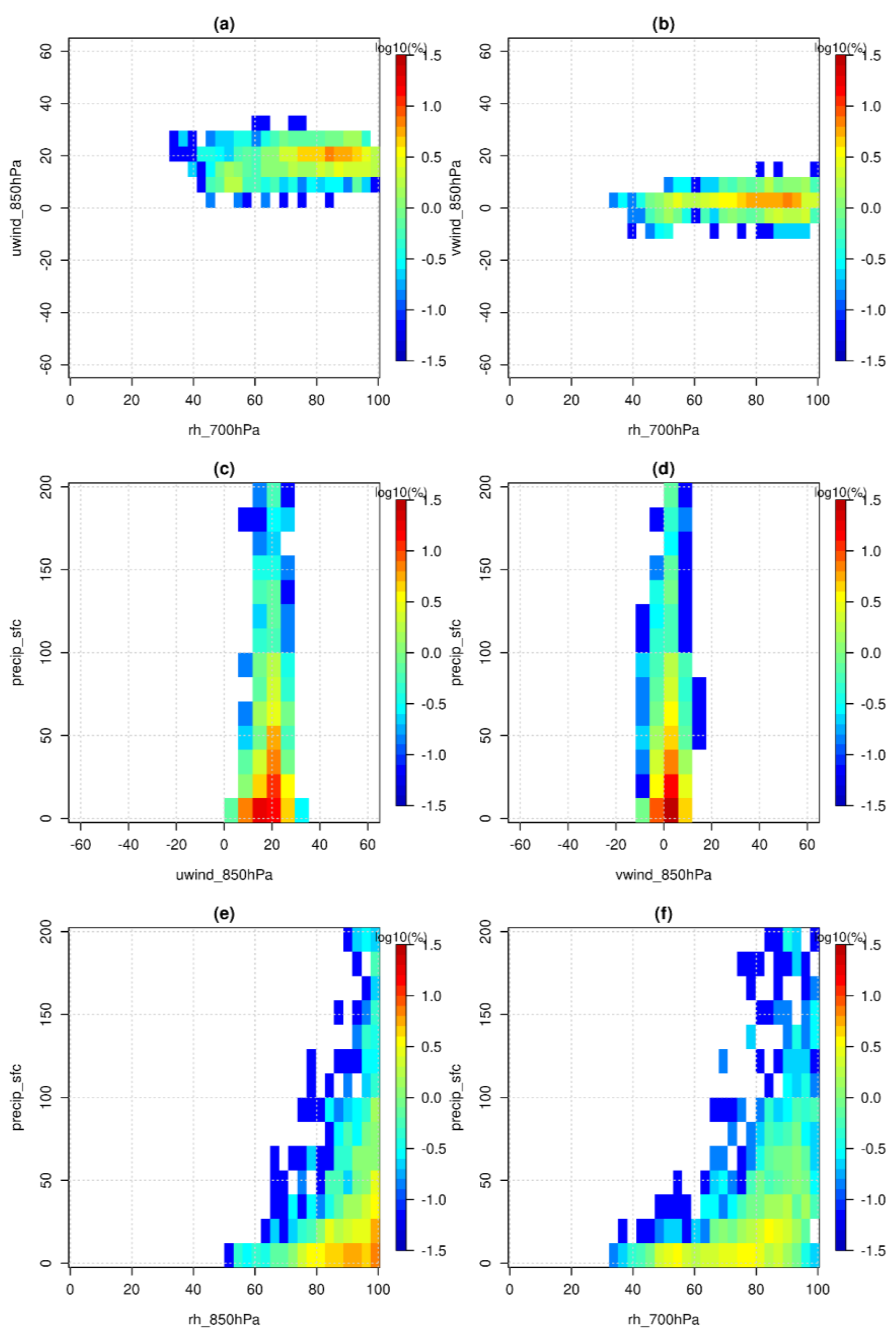

*Figure 2 Bivariate distributions between variables of interest for the purposes of threshold setting. Distributions derived from the period 1—11 August 2019 over the area indicated Figure 1. (a) 24h precipitation with UWIND@850hPa, (b) 24h precipitation with VWIND@850hPa, (c) UWIND@850hPa with relative humidity RH@700 hPa, (d) VWIND@850hPa with RH@700hPa, (e) RH@850 hPa with 24h precipitation, and (f) RH@700hPa with 24h precipitation.*

MvMODE was run over a large domain (5—40°N and 65-100°E), producing objects for each of the component fields (as univariate fields) as well as the super objects which track the overlaps. For analysis, the domain-wide results were filtered to extract only the objects with centroids relevant to the coastal region bounded by the Arabian Sea and the Western Ghats (5—25°N and 68-80°E).

Figure 3 to 5 provide daily snapshots of MvMODE output for the analyses and day 5 forecasts valid at 00 UTC on 16 August 2018, 8 August 2019, and 4 August 2020 (the full sequence of maps is provided in supplementary material). These are individual days taken from each of the three seasons' event time windows. The super objects are shown top right, populated with the model forecast and GPM accumulations. The MvMODE component objects are also shown for each of the forecast fields and the corresponding analyses. The similarities in the univariate objects between the days from the three seasons is evident. They show the broad swath of moisture at 700 hPa which sits just slightly to the south of where the westerly jet core is at 850 hPa. The univariate precipitation forecast object picks out the Kerala coast but also has some rain further to the NE, and which is well captured. The super objects tend to trim the extremities of the univariate objects offshore, and remove much of the rain in the NE, suggesting that this rainfall is not associated with the specific combination of ingredients being examined for the coastal region.

With all the super objects are identified and all the attributes computed, the subsequent analysis proceeds as in the univariate case, i.e. finding matches between forecast and analysis super objects. The following questions are the posed and are addressed in the following section.

1) Do the super objects capture features of interest in the analyses?
2) Do the super analysis objects contain the precipitation peak?
3) How does the precipitation super object analysis compare to a univariate analysis of the precipitation objects on their own?
4) Does the day 5 forecast capture a similar event?

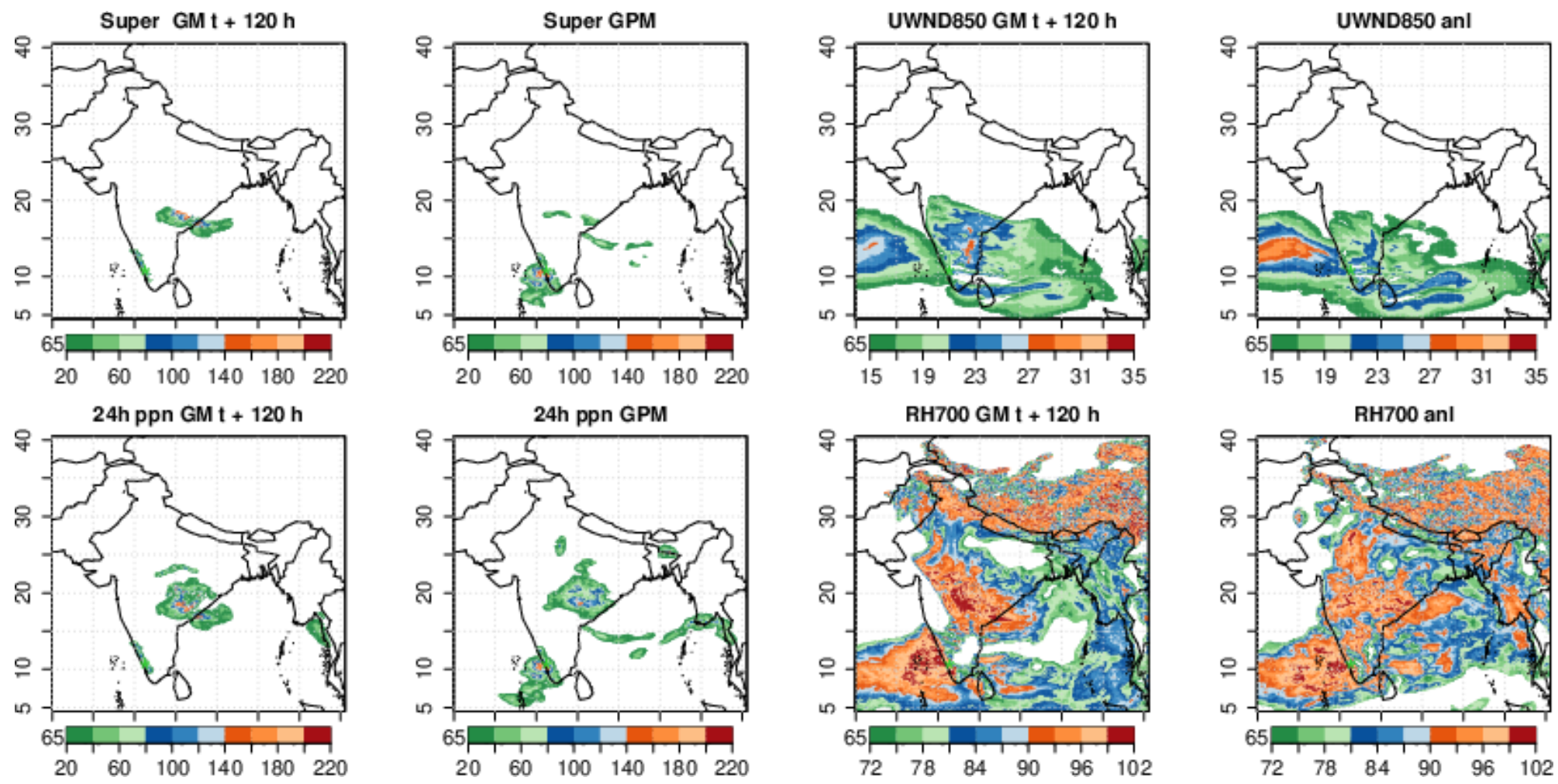

*Figure 3 Precipitation super objects and MvMODE component objects for the day 5 forecast valid at 00 UTC on August 16th, 2018, and the corresponding analyses. In the component fields the full-resolution forecast field is returned to the objects once they have been identified.*

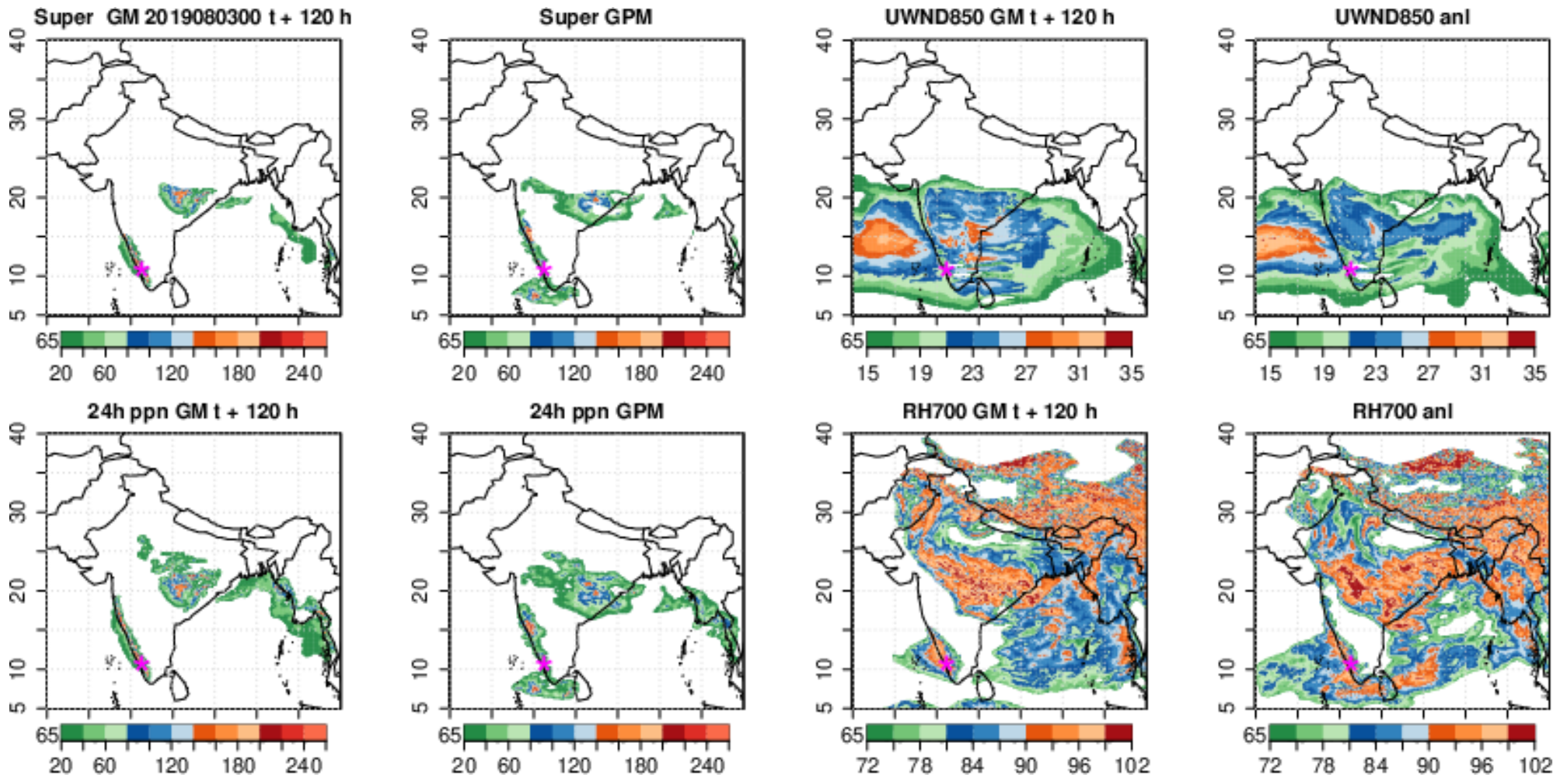

*Figure 4 Same as Figure 3 but for the day 5 forecast valid at 00 UTC on August 8th, 2019.*

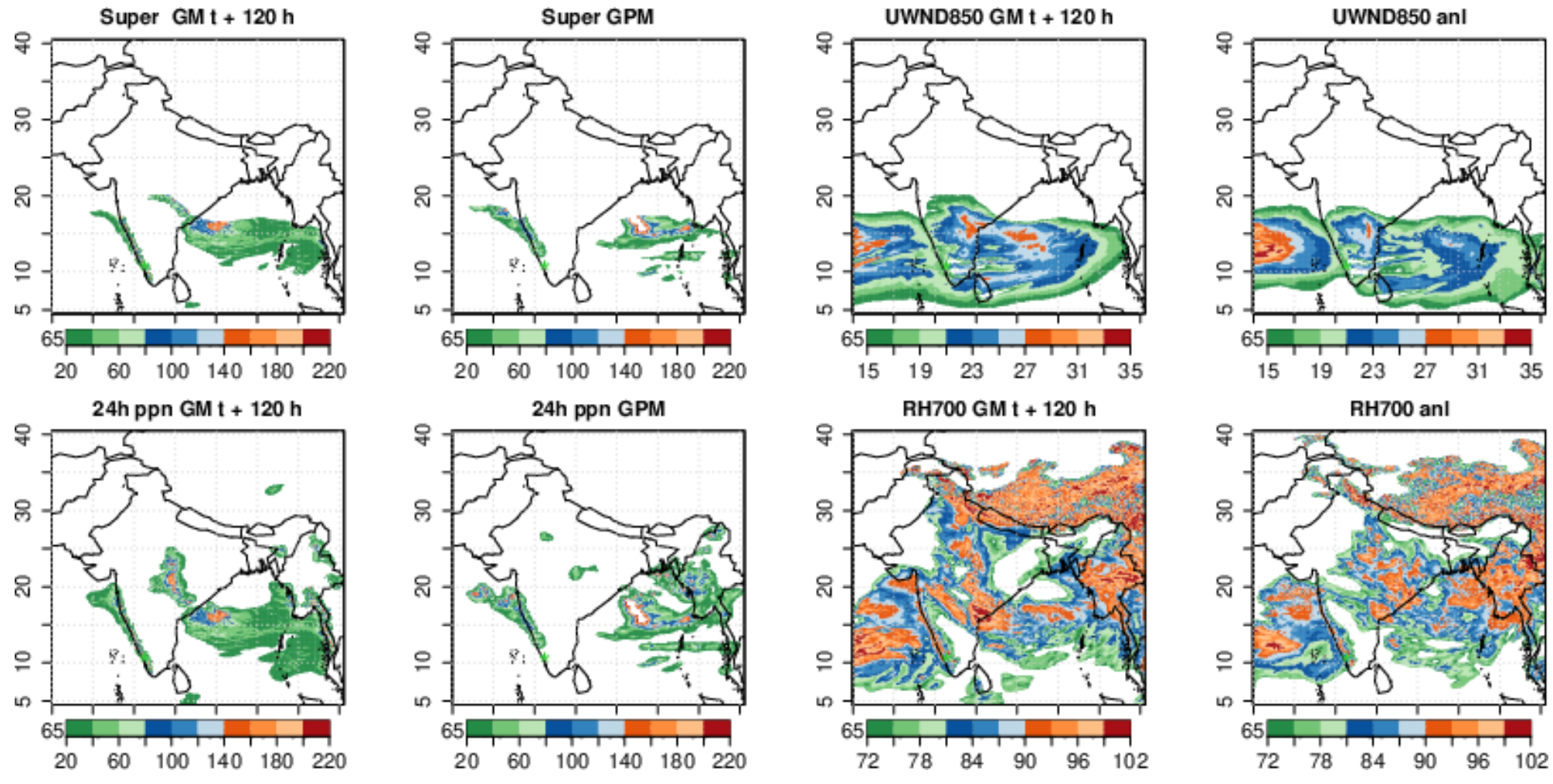

*Figure 5 Same as Figure 3 but for the day 5 forecast valid at 00 UTC on August 4th, 2020.*

## 4. Results

### *3.1 Single super object statistics*

Single super object statistics were compiled from the t+120h forecast objects and analysis (observed) objects in July/August of each year. The object centroids needed to be in the reduced region defined as 5—25 °N and 68—80 °E. Table 1 summaries the total number of objects found in the July-August period each year. It provides a summary of the proportion of the number and area of forecast and observed objects as well as the matches. When analysing the marginal distributions (i.e. forecast objects or observed objects only), it is usually more insightful to consider the single simple objects, and not the clusters, which add a layer of merging. Using single simple objects gives a better sense of the object size and an implicit insight into the degree of (de)fragmentation which may be inherent in the precipitation fields. It provides a better sense of whether the precipitation objects are large and contiguous, or small and plentiful, and how does this compare between the marginal distributions.

The total number of objects shows considerable inter-seasonal variation, as does the proportion of objects. During 2019 and 2020 there were more forecast objects than observed objects. The proportion of matched forecast objects also varies between 0.3 and 0.44. A larger proportion of observed single simple objects gets matched. This hints at the fact that matched observed cluster objects probably contain more single simple objects. More than half the total object area is contained within the forecast objects, suggesting that forecast objects are bigger. A good proportion (at least 75%) of this area gets matched. The area of matched observed objects is slightly lower but still at least 61%.

*Table 1 Summary of super objects over the three seasons for the day 5 super object forecasts and observed (analysis) objects imputed with the precipitation forecast field and GPM respectively for the reduced Kerala area (5—25 °N and 68—80 °E).*

| | **Jul/Aug 2018** | **Jul/Aug 2019** | **Jul/Aug 2020** |
|---|---|---|---|
| *Total number of single objects:* | 356 | 273 | 233 |
| *Total single forecast objects:* | 171 (0.48) | 148 (0.54) | 122 (0.52) |
| *Total matched single forecast objects:* | 56 (0.30) | 65 (0.44) | 41 (0.34) |
| *Total single observed objects:* | 185 (0.52) | 125 (0.46) | 111 (0.48) |
| *Total matched single observed objects:* | 71(0.38) | 61 (0.49) | 53 (0.48) |
| *Total area of objects (grid squares):* | 126826 | 140943 | 102846 |
| *Area of single forecast objects:* | 72375 (0.57) | 71443 (0.51) | 58530 (0.57) |
| *Area of matched single forecast objects:* | 54174 (0.75) | 57394 (0.80) | 43872 (0.75) |
| *Area of single observed objects:* | 54451(0.43) | 69500 (0.49) | 44316 (0.43) |
| *Area of matched single observed objects:* | 33299(0.61) | 52804 (0.76) | 32472 (0.73) |

The medians of key single simple forecast and observed object attributes are further summarised in Table 2. Also shown is the sample size for each group (forecast or observed) and year. Medians were preferred after a comparison of the means and the medians (not shown) showed that many of the attributions exhibited significant skewness. Under these circumstances the median is a better estimate of the underlying sample distribution. In terms of area, the results show considerable variability, also bearing in mind these are day 5 forecasts being considered. The largest anomaly is in the results for 2020 where there is a very large difference between the mean area of 315 and median of 31. For the other measures the conclusion would be that the forecast favours long and narrow objects (this will be explored further in Section 3.3), whereas the observed objects are still elongated but not necessarily to the same degree, yielding aspect ratios of 4 or 5-to-1 in the median, which given the coastline and terrain is not surprising. Finally, analysing the precipitation accumulations within the super objects, there is a discernible bias in the 90$^{th}$ percentile (not maximum) values, but the bias is not consistently in one direction between the years. Noting that these are the median 90$^{th}$ percentile accumulations from a global model with a notional resolution of ~12 km at these latitudes, these are substantial daily totals. The maximum 90$^{th}$ percentile super-object values are 103 (190), 633 (310), 278 (190) mm d$^{-1}$ for the forecast (GPM) for 2018, 2019, and 2020, respectively. As will be noted in the next section, the spot maximum values can be unrealistic, such that the 90$^{th}$ percentile values provide a much better guide.

*Table 2 Medians of day 5 key single simple super object attributes for each of the seasons.*

| **Attribute** | **2018** | | **2019** | | **2020** | |
|---|---|---|---|---|---|---|
| | **Fcsts (59)** | **Obs (71)** | **Fcsts (86)** | **Obs (96)** | **Fcsts (80)** | **Obs (85)** |
| *Area (grid squares)* | 287 | 139 | 291 | 374.5 | 31* | 244 |
| *Length (grid squares)* | 49.6 | 24.0 | 37.2 | 38.0 | 12.6 | 31.6 |
| *Width (grid squares)* | 10.7 | 9.4 | 13.9 | 17.4 | 4.3 | 13.8 |
| *90$^{th}$ percentile (mm d$^{-1}$)* | 82.9 | 64.2 | 75.8 | 91.0 | 48.0 | 70.4 |

*The mean in this instance was 315.

*4.2 Paired object statistics*

In this section the results for the matched pairs (between the analysis and the day 5 super objects of precipitation) are presented. Table 3 provides a summary of selected day 5 paired object attribute medians. Given these are day 5 forecasts, the centroid difference is less than ~15 grid square (~150 km), with the orientation error (angle difference) less than ~20 degrees, 50 % of the time. The orientation error may be driven by differences in orography and coastal definition in the model as well as by the difference in the centroid locations. The median area ratios, which represents a straight ratio of the forecast-to-observed object areas, is either remarkably good (for 2019) or skewed towards the forecast objects being much larger. Yet the overlap remains above 50%, which for a 5-day forecast suggests appreciable location predictability for a heavy precipitation event. The intensity ratios, which are based on the median intensity within the objects, shows a discrepancy in terms of median daily accumulations between the paired objects, providing evidence that the model and observed distributions do not have the same shape. The median interest scores for each of the seasons are identical, noting again that the minimum for matching is 0.7. Thus 50% of interest scores are 0.89 or better, but the interest scores for all matched pairs are 0.7 at worst.

*Table 3 Medians of selected day 5 paired object attributes for each of the seasons, with the number of paired clusters in brackets. Both the mean and median (in brackets) attribute values are provided.*

| **Attribute** | **2018 (52)** | **2019 (62)** | **2020 (37)** |
|---|---|---|---|
| *Centroid distance (grid squares)* | 17.06 | 14.83 | 14.3 |
| *Angle difference (degrees)* | 20.05 | 16.85 | 26.51 |
| *Area ratio* | 1.73 | 0.98 | 1.15 |
| *Intersection-over-area ratio* | 0.53 | 0.53 | 0.53 |
| *Percentile intensity ratio* | 0.70 | 0.74 | 0.79 |
| *Interest score* | 0.89 | 0.89 | 0.89 |

### *4.3 Compound event predictability*

Here the time series of the total areas on the days, during July and August, when there were identified matched pairs in each of the 2018, 2019 and 2020 seasons is considered. Figure 6 shows the multivariate super objects of precipitation in (a) and the univariate precipitation objects in (b) to show what impact the multivariate approach has. Recall also that these results are based on the day 5 (t+120h) forecasts, i.e. the focus is not only on whether MvMODE super objects can correctly identify the events in the analyses (GPM), but on what predictability exists in the day 5 super objects, i.e. whether the observed objects have a match. Thus, the paired objects provide both, since it checks that the event is identified in the analyses, as well as whether the day 5 forecast produced a match. Figure 6(a) shows the total area of all the matched super object pairs. Crucially, there were no unmatched objects. The bars track the total area encompassed by the matched objects (in grid squares). Given the filtering that is imposed on the object centroids, it can happen that only one object which is part of a matched pair has a centroid within Western Ghats sub-region. The bars are annotated with this information. There are six days where such a mismatch occurs (due to the filtering of results), of which there are three where the forecast object centroid of the matched pair is outside the sub-region (annotated with 0).

The focus in this study is on whether a multivariate methodology like MvMODE can identify multi-day events or episodes, here defined as at least 3 consecutive days in length, which were subsequently associated with high-impact consequences such as flooding. This is because many flood events are compound events, where accumulations build up over multiple days. During 2018 three episodes of three days or more were identified in both the GPM *and* the day 5 forecasts, with a 4-day episode coincident with the flooding period of 14—17 August. In 2019, the first three-day episode was in late July. This was then followed by a 9-day episode coincident with the flooding event. During 2020 there were three episodes of three consecutive days or more. The 7-day episode between 2—8 August 2020 coincides with the observed impacts. Two subsequent multi-day episodes followed later in August 2020, which were, as far as we can establish, not associated with the same level of impacts. Nevertheless, the 5-day forecast correctly identified the observed events. To reiterate, the events identified in July (and August) across the three seasons outside of the flooding event time windows are a product of the same synoptic drivers, but we know these events were not impactful in the same way. Area is therefore not a direct link to the local intensity or rain volume of an event. These elements need to be considered separately.

By contrast, the matched univariate MODE objects of rainfall exceeding 25 mm $d^{-1}$ shown in Figure 6(b), filtered over the same sub-region (and which represents one of the three inputs to MvMODE). As for the super-objects, there were no unmatched univariate objects, suggesting good predictability for day 5. The abrupt onset of the monsoon rains in this part of India in early July is more evident in the univariate objects. After the onset, the univariate matched pairs just keep appearing on a near daily basis. The exception is 2018 which has a distinct hiatus which lasts for close to 2 weeks. Nevertheless, it is hard to determine, based on this sequence, which of these days may be contributing to the observed flood-producing events, as the time series does not discriminate based on the mechanisms responsible for the rainfall, which is what we want to do. As Figure 6(a) shows, the multivariate combination of the UWIND@850, RH@700 and precipitation *can* provide a subset of these days, identifying a distinct and repeatable synoptic setup across the three years within the broader monsoon pattern, and which we know did cause flooding. Thus, we can conclude that the multivariate methodology has correctly identified the ingredients for the three flooding events during 2018, 2019 and 2020, though it ought to be pointed out that MvMODE identifies other episodes which were not directly linked to the occurrence of severe flooding, e.g. the July episode in 2018. This is because there are a multitude of factors contributing to a flood event occurring, and the rain volume is but one of them. The predictability in the day 5 forecast remains, given the presence of matched pairs. The general predictability demonstrated by these global forecasts is remarkable given the lead time.

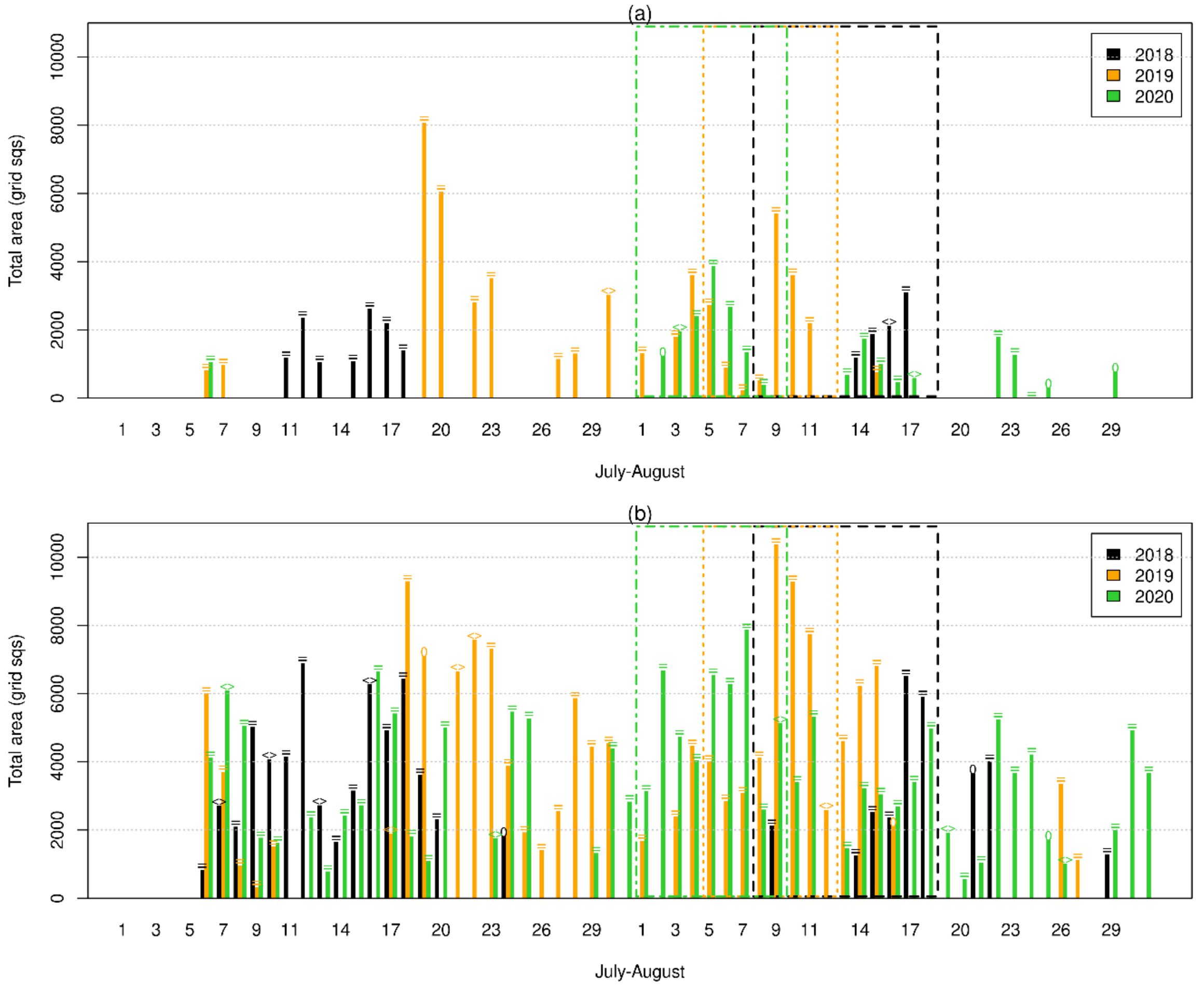

*Figure 6 (a) Total area of matched and filtered super objects per day for July and August for 2018, 2019 and 2020 for the Western Ghats sub-region. (b) The same but for univariate MODE precipitation objects of rainfall exceeding 25 mm d*$^{-1}$ *only. The flooding events were between 14—17 August 2018, 3—11 August 2019 and 2—8 August 2020. Main event windows are overplotted. Note that matched pairs can have one object that has a centroid outside the sub-region. This is reflected in the annotations. Annotations: =*

*indicates that both objects in the matched pair are in the sub-region; <> means that not all objects making up the matched pairs are in the sub-region; 0 = the forecast object of the matched pair falls outside the sub-region.*

Figure 7 provides the object-based event composites for the MvMODE components and precipitation super objects. Recall that the original fields are returned to the objects and either accumulated (for precipitation) or averaged. Both the analysis fields and the day 5 (t+120h) forecasts are shown. The super objects are populated with the precipitation fields before being accumulated.

There are several conclusions to draw from these plots. Firstly, the UWIND@850 fields show the low-level westerly jet, which is well captured in the day 5 forecasts. The maps show an apparent deceleration of the jet approaching the coast. This jet is well documented in the literature (Findlater, 1969; Fletcher et al., 2020; Joseph & Raman, 1966; Sunilkumar et al., 2024; Thapliyal, 2023), and vindicates the choice of UWIND@850 as input. Another interesting aspect seen in all the cases is the funnelling of the wind through a gap in the Western Ghats east of Kavalappara (the magenta asterisk on the maps), which is one of the locations that impacts were seen during the 2019 event (https://blogs.agu.org/landslideblog/2019/08/21/kavalappara-landslide-1/). It provides a clear indication of the barrier that the Western Ghats creates, which the low-level jet must ascend over, except where such gaps exist.

Secondly, all three episodes show a distinct reservoir of moisture over the region, which sits slightly offshore but straddles the coast, providing a plentiful source of moisture for orographic enhancement.

Thirdly, the univariate forecast precipitation objects demonstrate how the predicted precipitation is tightly locked to the model orography, leading to a narrow and extremely (unrealistically) intense region of rainfall, which does not span the region bounded by the terrain and the coast. This is a key model characteristic. For the 2019 and 2020 events the maximum forecast precipitation was 3224 and 3304 mm respectively! For 2018 the forecast event maximum was much more modest 520 mm. The GPM precipitation objects show broader west-east regions of rain, which stretch to the coast, and slightly offshore. This is reflected in the object statistics. The GPM event totals were 382, 816 and 1108 mm for the three years.

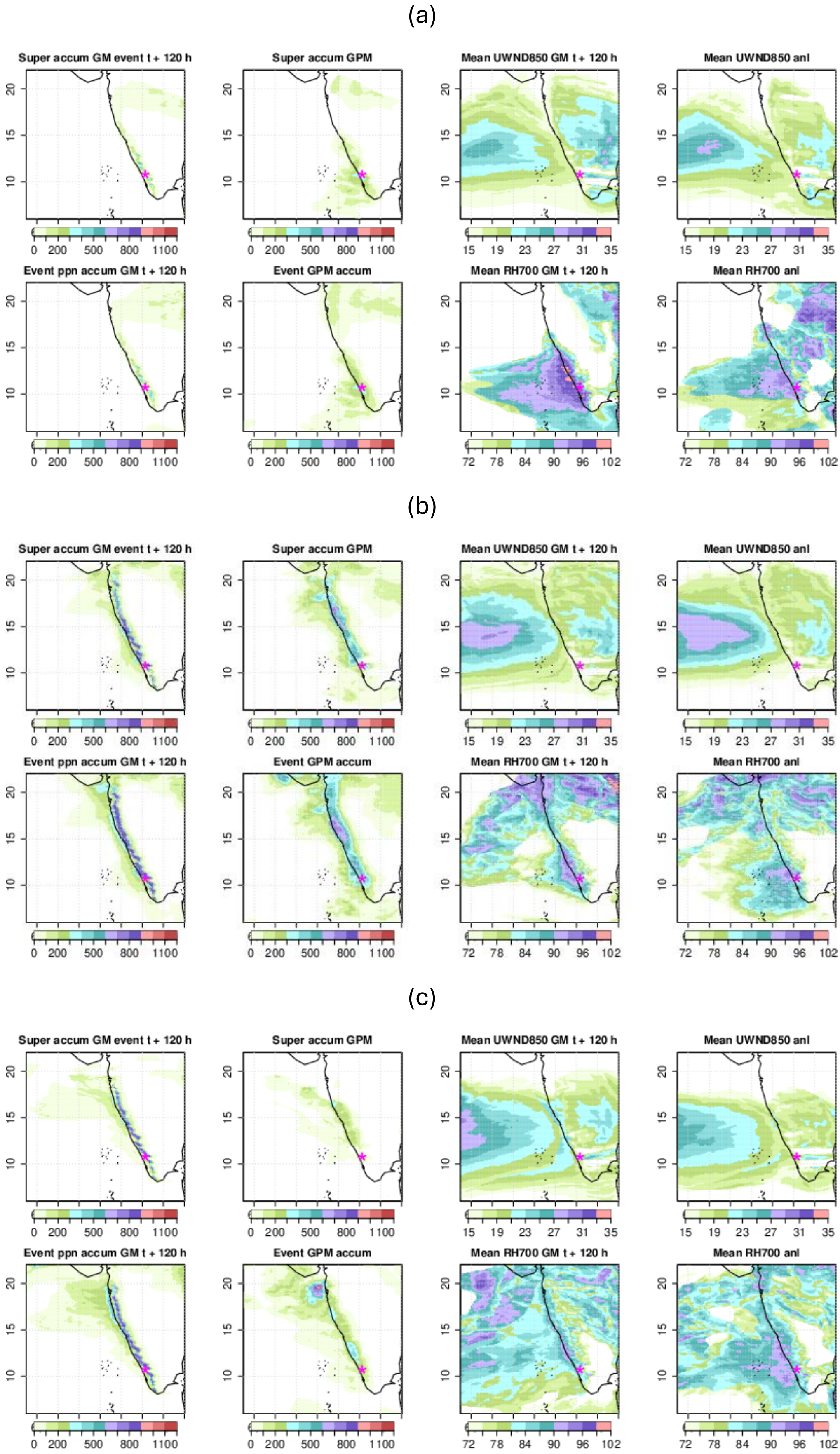

*Figure 7 Object-based event composites: accumulations (for precipitation) and mean object fields for (a) 2018, (b) 2019 and (c) 2020 events.*

Finally, some comments on the super objects. As alluded to elsewhere, they represent a subset of the univariate precipitation objects, focusing on the intersection of the UWIND@850, RH@700 and PPN24 objects. Many of the features of the univariate precipitation objects are replicated. Furthermore, the maxima found for the univariate precipitation objects were also the maxima of the super objects with one exception. The super forecast event maximum for the 2020 event was 2530 mm instead of 3304 mm, and the super observed event maximum was 393 mm instead of 1108 mm, and that is because the super objects exclude the rainfall associated with the offshore disturbance west of Mumbai, which MvMODE correctly removes because this event is not because of the interaction between UWIND@850 and RH@700.

## 5. Discussion and conclusion

The conclusion from the univariate precipitation-only analysis is that other mechanisms for precipitation exist, but using the multivariate approach identifies the events associated with a specific combination of atmospheric ingredients. The analysis shows that the days identified using this methodology occur in sequences or episodes which overlap or coincide with the flooding events during the 2018, 2019 and 2020 seasons. The combinations are repeatable, as demonstrated by the similarity in the synoptic setup for all three years and may provide a means for identifying future such events into the early medium range and perhaps beyond.

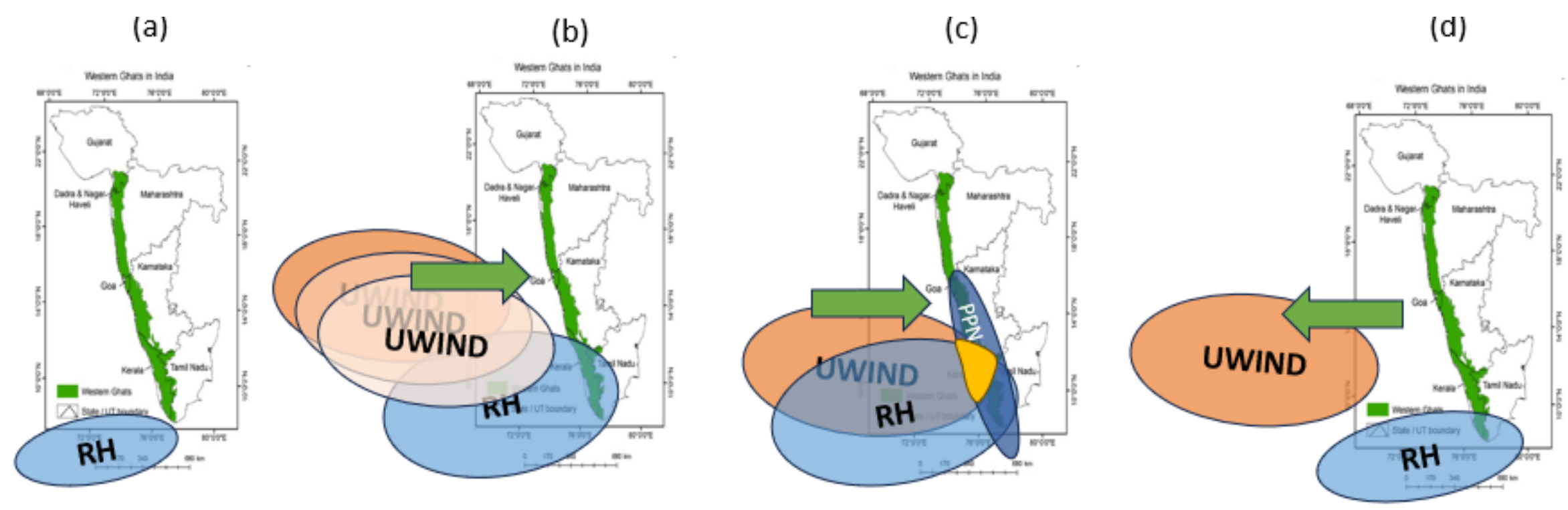

*Figure 8 Proposal of a conceptual evolutionary cycle for a quasi-atmospheric river.*

Given this repeatability, Figure 8 proposes an evolutionary cycle for these quasi-atmospheric river multi-day episodes. In (a) there is no sign of a low-level westerly jet. At the same time a reservoir of moisture builds to the south. Typically, there is little precipitation at this stage. In (b) there is a notable eastward progression and strengthening of the low-level, the leading edge of which is getting closer to the coast. Moisture continues to build and grow (in extent). At this stage there is still little engagement by the jet with the moisture source, resulting in little precipitation. By stage (c) the persistent low-level westerly jet decelerates as it approaches the coast and begins tapping into the moisture source, dragging it inland. This moist low-level flow is below the height of the terrain and is forced to rise, providing enhancement. In (d) the westerly jet retreats and/or weakens. The moisture has been depleted or shifted, and the precipitation stops. As noted earlier, the mechanism described in Figure 8 is not a true atmospheric river, but to a first order, behave very much like one. The low-level jet is necessary to tap into the reservoir of moisture which builds to the south and west of India as the monsoon progresses north-westwards.

To summarise, the questions posed in Section 3 are revisited.

- The super objects capture features of interest in the analyses.
- The super analysis objects contain the precipitation peak.
- The precipitation super objects are a subset (in number and extent) of the univariate precipitation objects, identifying only those rainfall events associated with a specific combination of synoptic drivers.
- In all cases, the GPM precipitation super objects are matched by day 5 precipitation super objects, with location errors of less than ~150 km for 50% of the matches. This suggests useful predictability and the potential for identifying future compound events into the early medium-range, if not beyond. Future work will focus on exploring the predictability at even longer ranges.

Several caveats with respect to the outcomes of this study should be mentioned. a) The day 5 results, with respect to predictability, are model dependent and not necessarily transferable to another model, though the existence of the synoptic drivers in respective analyses and observations is robust. b) The event rainfall totals are dependent on the dataset used and GPM is not without its issues. As a bulk indicator of event magnitude, GPM should give a good enough steer. c) As disclosed, the GM produces unnatural looking and occasionally unrealistic amounts of rainfall, which is linked to the resolution, the parameterised convection, and the model orography. The nature of the forecast precipitation fields is therefore dependent on the resolution, the representation of orography and whether the convection is parameterised or not. It is likely that the forecast precipitation fields would need to be treated differently for higher spatial resolution models.

This study shows that a multivariate feature-based approach can provide a much more nuanced and targeted methodology, whether performing routine verification, event-based studies or understanding model systematic behaviour.

**Acknowledgements**

I would like to thank Tracy Hertneky from the Developmental Testbed Center (DTC, NCAR) for her help in getting the prototype version of MvMODE in MET11.0.0 working.

This work and its contributors (Marion Mittermaier) were funded by the Met Office Weather and Climate Science for Service Partnership (WCSSP) India project, which is supported by the UK Department for Science, Innovation & Technology (DSIT). WCSSP India is a collaborative initiative between the Met Office and the Indian Ministry of Earth Sciences (MoES).

**Availability statement**

Operational global forecasts are archived on internal Met Office storage systems and are subject to standard Met Office retention policy. A standard freedom of information request can be used to access the forecast data used.

The IMERG GPM (https://pmm.nasa.gov/data-access/downloads/gpm) Final Precipitation version 6 (V06) level 3 product was provided by the NASA/Goddard Space Flight Center's and PPS, which develop and compute the dataset as a contribution to GPM project, and archived at the NASA GES DISC.

MET and METplus can be downloaded from https://github.com/dtcenter.